\documentclass{aa}  
\usepackage{graphicx}
\usepackage{epsfig}
\usepackage{txfonts}
\begin{document}
   \title{Probing IGM large-scale flows: warps in galaxies at
   shells of voids}

   \subtitle{}

   \author{M. L\'opez-Corredoira\inst{1}, E. Florido\inst{2},
   J. Betancort-Rijo\inst{1,3}, I. Trujillo\inst{1,4}, C. Carretero\inst{1,5},  
   A. Guijarro\inst{2,6}, E. Battaner\inst{2}, S. Patiri\inst{1,7}}

   \offprints{martinlc@iac.es}

\institute{
$^1$ Instituto de Astrof\'\i sica de Canarias, C/.V\'\i a L\'actea, s/n,
E-38200 La Laguna (S/C de Tenerife), Spain\\
$^2$ Departamento de F\'\i sica Te\'orica y del Cosmos, 
Universidad de Granada, Spain\\
$^3$ Departamento de Astrof\'\i sica, Universidad de La Laguna, Tenerife,
Spain\\
$^4$ School of Physics and Astronomy, University of Nottingham, 
University Park, Nottingham NG7 2RD, U.K.\\
$^5$ Estin \& Co Strategy Consulting, 43 Av. de
Friedland, 75008 Paris, France\\ 
$^6$ Centro Astron\'omico Hispano Alem\'an, Almer\'\i a, Spain\\
$^7$ Case Western Reserve University, Cleveland (Ohio), USA}

   \date{Received xxxx; accepted xxxx}

 
  \abstract
  {Hydrodynamical cosmological simulations predict flows of the intergalactic medium along 
the radial vector of the voids,
approximately in the direction of the infall of matter at the 
early stages of the galaxy formation.}      
   {These flows might be detected by analysing the dependence of
   the warp amplitude on the inclination of the galaxies at the shells
   of the voids with respect to the radial vector of the voids.
   This analysis will be the topic of this paper.}
   {We develop a statistical method of analysing the correlation
    of the amplitude of the warp and the inclination of the galaxy 
    at the void surface. This is applied to a sample of 
    97 edge-on galaxies from the Sloan Digital Sky Survey. Our results are
   compared with the theoretical expectations, which are also derived in
   this paper.}
   {Our results allow us to reject the null hypothesis
   (i.e., the non-correlation of the warp amplitude and the inclination
   of the galaxy with respect to the void surface) at 94.4\% C. L.,
   which is not conclusive. The absence of the radial flows cannot be 
   excluded at present, although we can put a constraint on the maximum 
   average density of baryonic matter of the radial flows of 
   $\langle \rho _b\rangle <\sim 4\Omega _b\rho _{\rm crit}$.}
   {}

   \keywords{intergalactic medium --- galaxies: statistics ---
   galaxies: kinematic and dynamics --- large-scale structure of Universe}
\titlerunning{Warp / voids}
\authorrunning{L\'opez-Corredoira et al.}

   \maketitle
%

\section{Introduction}

Warps seem to be an almost universal structural feature in spiral galaxies. 
Indeed, most of the spiral galaxies for which we have relevant information on
their structure (because they are edge on and nearby) present warps
in their stellar and gas distributions.
S\'anchez--Saavedra et al. (1990, 2003) and Reshetnikov \& Combes (1998) 
show that nearly half of the spiral galaxies of selected samples 
are warped, and many of the rest
might also be warped since warps in galaxies with low inclination are difficult
to detect. They are more clearly observed in the HI distribution
(see e.g. van der Kruit 2007 and references therein).
Warps are also detected at about $z=1$, even with a larger amplitude
(Reshetnikov et al. 2002). 

Despite the compelling observational evidence of warps in the
spiral discs, there is consensus on what could be the origin of this
property of the galaxies. Nevertheless, it seems clear that warps should be
produced by an interaction of the disc with an external element. In fact, Hunter
\& Toomre (1969) showed that in an isolated galaxy (without a dark matter halo),
an initial warp would soon disappear and leave as its only trace a thickening
of the edge of the disc.

The number of ideas suggested to explain the origin of the warp in discs is
vast. One explanation for the warps is
gravitational tidal effects due to satellite galaxies. At least in the Milky
Way galaxy, this explanation does not work with Magellanic Clouds
as satellite (Hunter \& Toomre 1969), and it is controversial
whether it works in combination with the amplification of the halo
(as proposed by Weinberg 1998 and criticised by 
Garc\'{\i}a--Ruiz et al. 2002). Also, the
intergalactic magnetic field has been suggested as the cause of galactic warps
(Battaner et al. 1990; Battaner et al. 1991; Battaner 
\& Jim\'enez--Vicente 1998).

Following the evidence that galaxies seem to be embedded in a massive
dark matter halo, the interaction between the halo and the disc was explored. 
Ideas like `dynamical friction' between the disc and a spherical halo (Bertin \&
Mark 1980; Nelson \& Tremaine 1995), a flattened halo misaligned with the disc
(Toomre 1983; Dekel \& Shlosman 1983; Sparke \& Casertano 1988; Kuijken 1991),
or resonant interactions with a triaxial halo (Binney 1981) were explored. All
these ideas, however, were rejected when the dark matter halo was modelled
correctly as a deformable mass of collisionless particles, rather than as a
rigid body (Binney et al. 1998). 
Since a warp represents a misalignment of the disc's inner and outer angular
momentum, Ostriker \& Binney (1989) and Jiang \& Binney (1999) 
proposed a model in which warps are generated through accretion of
material into the halo with a misaligned spin that changes the major axis
of the halo with respect to the disc and consequently produces a torque
over the disc. 
There is a need for substantial accretion of low angular momentum 
material from the IGM into the galaxies (Fraternali et al. 2007), and the
direction of the net angular-momentum vector of the material that is currently
being accreted should be constantly changing (Quinn \& Binney 1992). 

Also based on infalling of material, but with a much weaker dependence 
on halo properties, some works (Mayor \& Vigroux 1981; 
Revaz \& Pfenninger 2001; L\'opez-Corredoira et al. 2002; 
S\'anchez--Salcedo 2006) have proposed a mechanism
for the formation of the warp in terms of 
the infall of a very low density intergalactic
medium onto the disc without the dynamical intervention of an intermediate halo.
Both S-type and U-type warps can be produced by this interaction 
(L\'opez-Corredoira et al. 2002; Saha \& Jog 2005).
Even if there are other mechanisms able to produce warps, at least
we know that the infall of material onto the disc 
will always produce warps.

If the infall of material is relevant to the formation of the warp of the disc,
the orientation of the galaxies within the cosmological large--scale structure
where they are embedded should have an effect on the formation of these
features. There is growing evidence that disc galaxies are not oriented 
randomly,
but their angular momentum primarily point parallel to the filaments
(or sheets) where they are located. In the supergalactic plane, there is  a hint
of an excess of galaxies whose angular momentum lie in this plane (Kashikawa \&
Okamura 1992; Navarro et al. 2004). Beyond the local universe, Trujillo et al.
(2006) show at the 99.7\% level that spiral galaxies located on the shells of
the largest cosmic voids ($r>10\ h^{-1}$Mpc) have rotation axes that lie 
primarily on the void
surface. Paz et al. (2008) point out that the angular momentum of flattened 
spheroidals in SDSS galaxies tends to be perpendicular to the large-scale 
structure. These alignments are expected to be a consequence of the gain in
angular momentum of the galaxies at the early stages of their formation, when
both the baryonic component and the dark matter protohalo are suffering tidal
torques from neighbouring fluctuations. Using N--body simulations, the
alignments of the angular momentum of the haloes with the large--scale
distribution have been also found (Porciani et al. 2002; Bailin \& Steinmetz
2005; Brunino et al. 2007; Arag\'on-Calvo et al. 2007; Hahn et al. 2007;
Paz et al. 2008).

The aim of this paper is to check whether the orientation of the spiral galaxies
in the void surfaces is related to the presence of a warp or not. In contrast to
filaments (which are strongly affected by redshift--space distortion), large
cosmological voids are a feature easy to characterise from the observational
point of view. In addition, another important advantage of the void scheme is that
(because of the radial growing of the voids) the vector joining the centre of
the void with the galaxy position is a good approximation of the direction of
the maximum compression of the large--scale structure at that point.
Consequently, the radial vector of the void at the galaxy position 
approximately represents the direction of the infall of matter at the 
early stages of the
galaxy formation. At later epochs, however, most of the accretion of material in
the galaxy is expected to be through the filaments (i.e. parallel to the void
surface). According to L\'opez-Corredoira et al. (2002), the infall of material
should produce a correlation between the orientation of the galaxy and the
amplitude and direction of the S-component, or the U-component or both of them. 
We want to check this hypothesis here. The aim of this paper is producing
a method for analysing the relationship of the warp amplitude in galaxies
with the inclination
of the galaxy with respect to the line ``centre of void''-galaxy (to check the
early accretion of material). This method is then applied to 
the edge-on galaxies and void catalogue 
used by Trujillo et al. (2006)
for available images from Sloan Digital Sky Survey (SDSS) survey. The 
work presented here is an attempt to
observationally characterise the influence of the large--scale structure (and,
consequently, the cosmic infall of material) on the formation of the warps.  
Some works have previously dealt with no random orientations 
of warps on large scales (Battaner et al. 1991) 
or in the Local Group (Zurita \& Battaner 1997), but not 
at the void shells.

\section{Definitions}

\subsection{Warp amplitude}

\begin{figure}
\vspace{1cm}
{\par\centering \resizebox*{10cm}{10cm}{\includegraphics{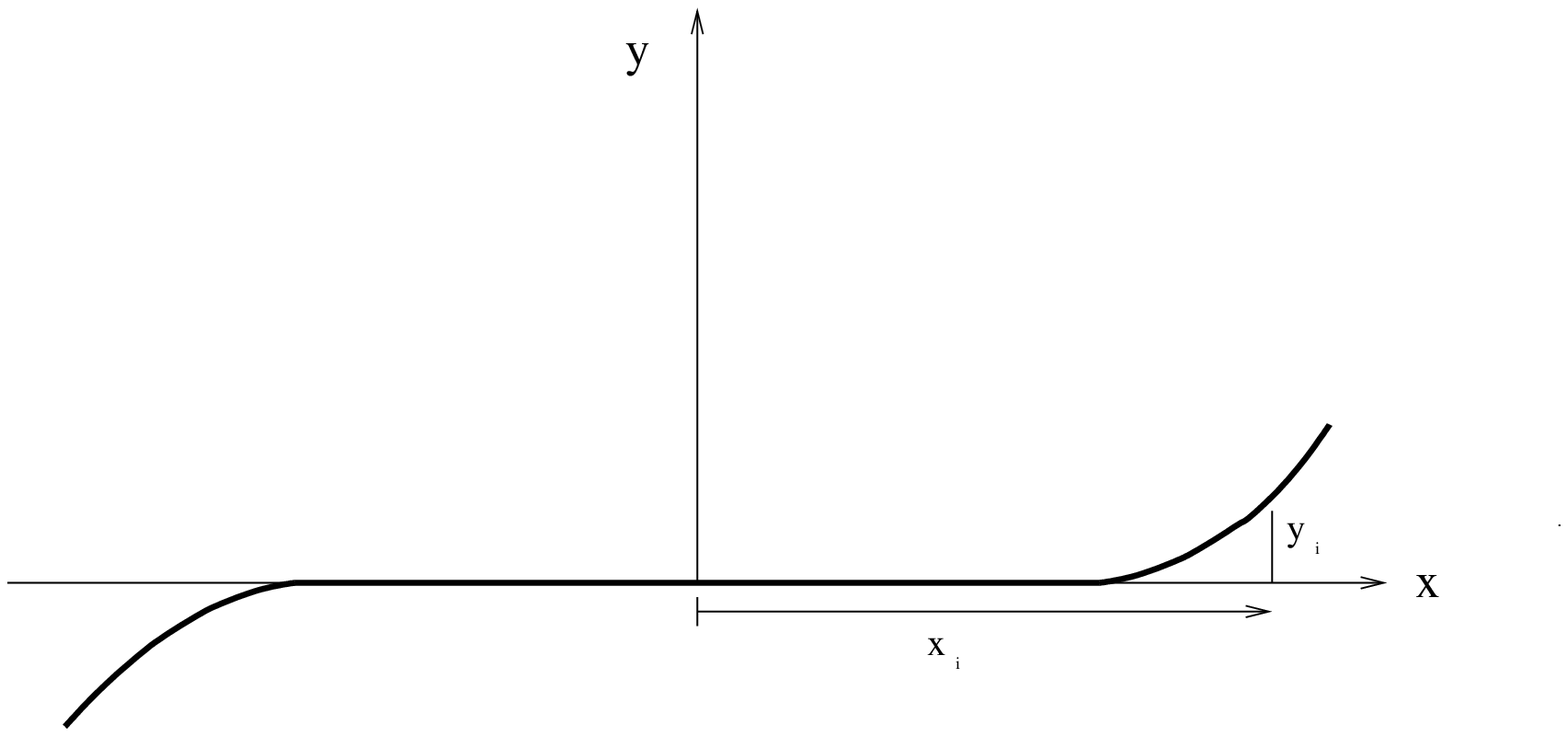}}}
\caption{Graphical representation of the warp measurement.}
\label{Fig:florido}
\end{figure}

To define a warp amplitude, we first rotate the galaxy to have
the mean plane of the galaxy coincident with the constant declination
axis in the local plane of the sky (perpendicular to the line of sight).
The position angle is calculated with an iterative method that fits the
central part of galaxies (size of galaxy/2) to a straight line. This
method uses the position angle from Trujillo et
al. (2006) as starting point. 
The position angle was determined in this way to an
accuracy of about 0.5 degrees. This error was adopted like that of the RMS
using the mean least square method in the rotation procedure. We then
have a right and a left part of the galaxy, each having its own warp.
The right part is the one with a lower right ascension. 
To quantitatively estimate the warp amplitude we define the
warp parameter $W$, on the left(l) or right(r) side of the galaxy, as

\begin{equation}
W_{(r\ or\ l)}= \frac{\int _0^{L_{(r\ or\ l)}} x\ y\ dx}
{(\int _{-L_l} ^{L_r}dx)^3}
\label{formulaW}
,\end{equation}
where $L_{(r\ or\ l)}$ is the radius (left or right) of the disc within
the limits in which the disc is visible $\sigma \equiv signal/noise>3$;
$\sigma$ =$\sqrt{\sigma^{2}_{\rm{sys}} + \sigma^{2}_{\rm{std}}}$, where
$\sigma_{\rm{sys}}$ is the systematic error and
$\sigma_{\rm{std}}$ the standard deviation) 
and $y$ is the height of the disc at position of pixel $x_i$,
being $x\ge 0$ (Fig. \ref{Fig:florido}).
The y-values are obtained as the peaks of
Gaussians fits in the light distribution perpendicular to the plane. As said,
we only considered data with intensity greater than 3$\sigma$. 
An example of the result of our analysis
is presented in Fig.\ref{Fig:sample}, where the warp curve is drawn only for
those values with an error bar less than 0.5$\arcsec$. This estimated
error bar can be computed by scaling the standard deviation (1$\sigma $
error) by the measured chi-squared value. Then, $W_r$ will be positive for
warp towards increasing declination, and vice versa for $W_l$. 
A large warp can reach values of $W=5\times 10^{-3}$ 
and a barely perceptible warp $W=5\times 10^{-4}$.

Expression (\ref{formulaW}) is adimensional, 
therefore the value of $W$ only depends on
the shape of the edge-on galaxy but not on the intrinsic size or on the
distance of a galaxy [neglecting the change of 
the factor $(1+z)^4$ (i.e. cosmological dimming)
in the surface brightness of the galaxies
throughout our sample since most of them are at a similar $z\sim 0.1$]. 
For numerical purposes and working in pixels, we use the discrete expression

\begin{equation}
W_{(r\ or\ l)}= \frac{\sum _0^N x_iy_i}{(L_l+L_r)^3}
,\end{equation} 
where $x_i\ge 0$ and $x_N=L_{(r\ or\ l)}$. 
The error in estimating $W$ is dominated by the imperfect rotation
step of the galaxy when it is rotated to make the major axis coincident
with the x axis. As mentioned before,
this rotation is performed with an error of 0.5
degrees (i.e. about 0.008 radians). 
This introduces an error in $W$ given by 

\begin{equation}
\Delta W_{(r\ or\ l)}=\frac{\int _0^{L_{(r\ or\ l)}} 
x^2\tan{0.008}\ dx}{(L_r+L_l)^3}\sim 3.3\times 10^{-4}
.\end{equation}
Further details of this method of warp measurement are
given in Guijarro et al. (2008).

In a S-shape warped galaxy, $W_r$ and $W_l$ have the same sign. In a 
U-shaped galaxy, $W_r$ and $W_l$ have different signs.
We define the variables $S$ and $U$ as
\begin{equation}
S\equiv W_r+W_l
,\end{equation}
\begin{equation}
U\equiv W_r-W_l
\label{defU}
.\end{equation}
If the warp is of the type with integral-sign [S-warp, see Fig. 
\ref{Fig:warp}(left)], $S$ will be different from 
zero, positive or negative, and $U$ will be zero if it is perfectly
symmetrical or has a low value if there is some asymmetry. 
Otherwise, if the warp is predominantly cup-shaped [U-warp, see Fig.
\ref{Fig:warp}(right)], $U$ will be different from zero and 
$S$ zero or very low, since it expresses
the degree of asymmetry with respect to a perfect U-shape.
An L-warp will have $|S|\approx |U|$. 
The combination of S-warps and U-warps explain the asymmetry of the
warps (L\'opez-Corredoira et al. 2002; Saha \& Jog 2006) and
the values of $S$ and $U$ give us the degree of each component to
the total warp. We assign a value of
$S$ and $U$ to all the galaxies in our sample and
make statistics with these numbers, which quantify the S-component
and the U-component.

A serious difficulty arising in any observational study of warps is that
companions, spiral arms, 
and other effects may mimic warps. The error introduced by a
misidentification are difficult to 
evaluate. However, the images do not suggest that the
warps are confused with spiral arms. On the other
hand, the companions which are far away from the
plane of the main galaxy are not
confused with the warp, and if they were very close to the galactic
outskirts, the galaxy would be removed from our list.

\subsection{Inclination of the galaxy with respect to the
centre of the void}

\begin{figure*}
\vspace{1cm}
{\par\centering \resizebox*{6cm}{6cm}{\includegraphics{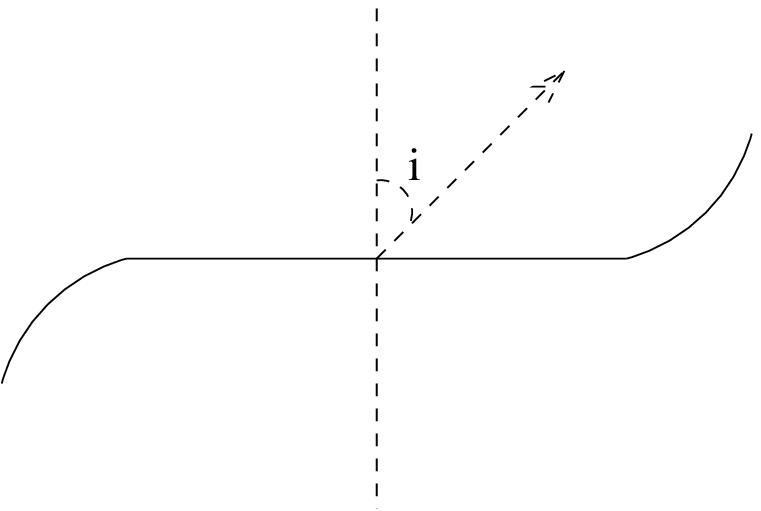}}
\hspace{1cm}\resizebox*{6cm}{6cm}{\includegraphics{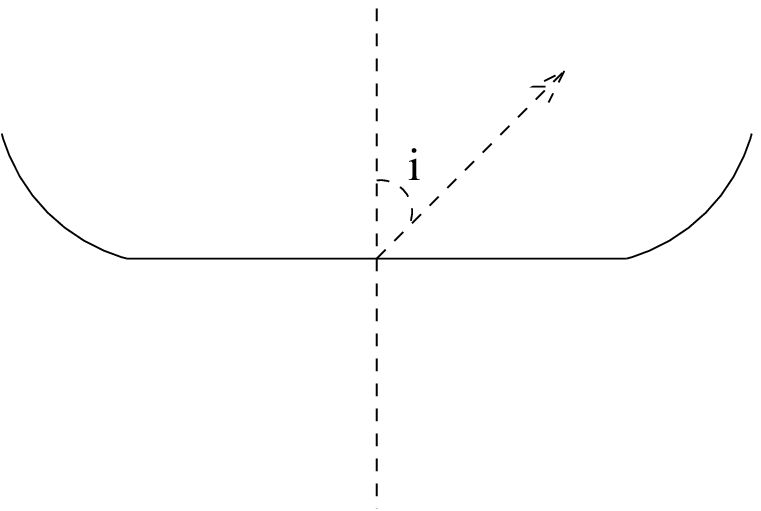}}\par}
\caption{Left: Graphical representation of a perfect S-warp ($S\ne 0$,
$U=0$). Right: Graphical representation of a perfect U-warp ($U\ne 0$, $S=0$). 
North (higher declination) is up, south is down. 
``i'' stands for the inclination between the line 
``centre of void''-galaxy and the rotation axis of the galaxy.}
\label{Fig:warp}
\end{figure*}

For each galaxy, given its position angle and the position with respect
to the centre of the void (see Trujillo et al. 2006 for details), 
we calculated the inclination of the rotation
axis with respect to the line ``centre of void''-galaxy. The sense of
the rotation axis makes the ``right'' warp positive,
that is, toward increasing declination. And the inclination $i$ is defined
positive (between 0 and $\pi $) if the line ``centre of void''-galaxy 
is to the right (decreasing position angle) of the rotation axis or
negative (between 0 and $-\pi $) otherwise. 
Figure \ref{Fig:warp} illustrates this.

The error in this inclination stems from the error on
the distance to the galaxies in Trujillo et al. (2006) sample.
Due to the intrinsic motion of the galaxies away from the Hubble flow,
this error is estimated to be around
4 $h^{-1}$Mpc, and the error in the distance to the centre of the void,
around 2 $h^{-1}$Mpc. Taking into account that the average distance
of the galaxies to the centre is $\approx 12$ $h^{-1}$Mpc.
This leads to an average error of $\approx 14^\circ $. Since these
errors are statistical and not systematic, they will not affect
the average signal that we find in the data, but will only
decrease the signal-to-noise ratio.

\section{Theoretical predictions from a model of warp formation
in terms of accretion of IGM onto the galactic disc}

\subsection{Warp dependence on inclination of the galaxy}

The predictions of the model with accretion of intergalactic
medium (IGM) onto the disc for an average Milky Way-like galaxy
is given in L\'opez-Corredoira et al. (2002, Fig. 11). A 
fit of the curves in that figure gives the theoretical values $S_t$ and
$U_t$:

\begin{description}

\item[S-component:]\ 

\begin{equation}
S_t(\theta [deg.],v; 0<\theta <90)\approx S_0(v)
\end{equation}\[\times
\sin (4.99-1.26\theta +0.164\theta ^2-0.00324\theta ^3
+1.99\times 10^{-5}\theta ^4),\]
\[S_t(\theta[deg.],v; 90<\theta <180)=-S_t(180-\theta ,v);
\]\[ S_t(\theta[deg.],v;
-180<\theta<0)=S_t(\theta+180,v)\]

\item[U-component:]\ 

\begin{equation}
U_t(\theta [deg.],v; 0<\theta <90)\approx U_0(v)\cos (1.16
\label{U0}
\end{equation}\[
-0.295\theta +0.0796\theta ^2-0.00173\theta ^3
+1.09\times 10^{-5}\theta ^4)
,\]
\[U_t(\theta [deg.],v; 90<\theta <180)=-U_t(180-\theta,v);
\]\[ U_t(\theta [deg.],v; 
-180<\theta <0)=U_t(-\theta ,v),\]
\end{description}
where $\theta $ in these expressions is the
direction of the IGM wind with respect to the rotation axis of the galaxy,
and $v$ the  relative velocity of the wind. Together, 
$U_0$ and $S_0$ represent the maximum
amplitude of the $S_t$ and $U_t$ that we calculate
in the following sections.

\begin{figure*}
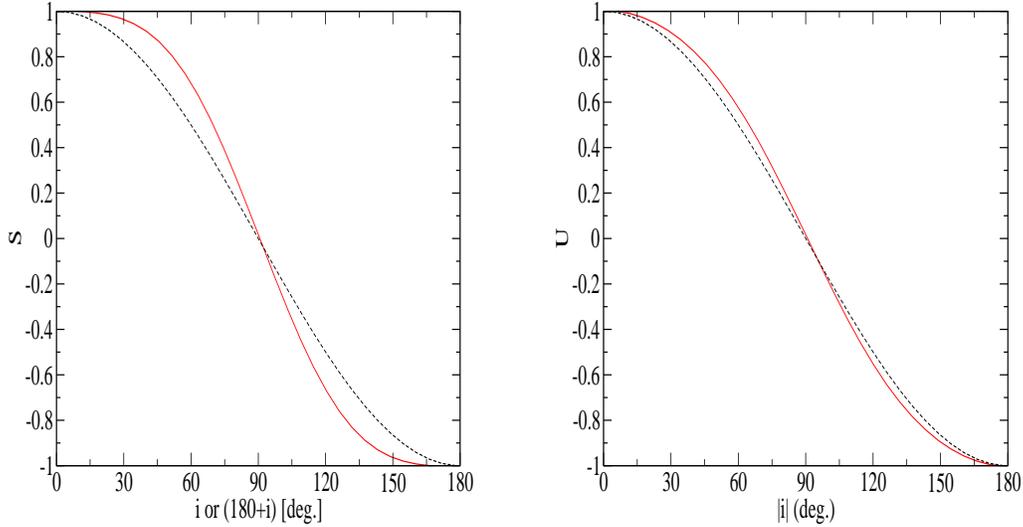

\vspace{1cm}
{\par\centering \resizebox*{6.2cm}{7cm}{\includegraphics{Steor.eps}}
\hspace{1cm}\resizebox*{6.2cm}{7cm}{\includegraphics{Uteor.eps}}\par}
\caption{Dependence of $\overline{S}$ and $\overline{U}$ on $i$ predicted by the
theory. Solid line: expected trend if the warps were 
produced by intergalactic
winds flowing radially outwards in the void, according to
expressions (\protect{\ref{S}}) and (\protect{\ref{U}}), 
normalized to a maximum height of one. Dashed line: $\cos i$.}
\label{Fig:teor}
\end{figure*}

Assuming there is a wind flowing radially outwards in the void with velocity
$\overline{v_1}=200$ km/s (details will be given in
Betancort-Rijo \& Trujillo 2008), we must add
the dispersion of velocities of the galaxies: $\sigma _1=215$ km/s, 
$\sigma_2=209$ km/s
(Betancort-Rijo \& Trujillo 2008) in 
the radial velocity $v_1$ (the projection of the velocity into the
radial direction of the void) and the perpendicular component $v_2$ with
respect to the radial direction of the void with angular azimuth $\phi $.
To obtain these numbers, Betancort-Rijo \& Trujillo used the
linear theory of growing fluctuations in the large-scale structure.
They computed the r.m.s. of the corresponding components of
the velocity of mass particles on the surface of a void of 10$h^{-1}$Mpc
with respect to its centre of mass. These numbers agree
within a few per cent with the numbers found in numerical simulations.
Hence, the average warps are given by

\begin{equation}
\overline{S _t} (i)=\frac{1}{\pi} \int _0^\pi d\phi \int _0^\infty dv_2
\int _{-\infty }^\infty dv_1P(v_1,v_2)S_t(\theta ,v)
\label{S}
,\end{equation}
\begin{equation}
\overline{U _t}(i)=\frac{1}{\pi} \int _0^\pi d\phi \int _0^\infty dv_2
\int _{-\infty }^\infty dv_1P(v_1,v_2)U_t(\theta ,v)
\label{U}
,\end{equation}
\begin{equation}
\theta =\cos ^{-1}\left (\frac{v_1\cos i +v_2\sin i\cos \phi }{v}\right)
\end{equation}
\begin{equation}
P(v_1,v_2)=\frac{1}{\sqrt{2\pi }\sigma _1}\frac{v_2}{\sigma _2^2}\exp{\left (
-\frac{(v_1-\overline{v_1})^2}{2\sigma _1^2}
-\frac{v_2^2}{2\sigma _2^2}\right)}
,\end{equation}
\begin{equation}
v=\sqrt{v_1^2+v_2^2}
,\end{equation}
where $S_t(\theta ,v)\propto v^2$, $U_t(\theta ,v)\propto v^2$ 
because the
warp amplitude is proportional, both in the S-shape and U-shape, to $v^2$ 
(L\'opez-Corredoira
et al. 2002; Eqs. (39), (45)). 
When we make all these calculations, we get
a dependence of $S_t(i)$ and $U_t(i)$ on $i$ 
which is close to $\cos i$ (see Fig. \ref{Fig:teor}), although closer in the case of $U_t(i)$
than in the case of $S_t(i)$.
Therefore, from now onwards, we will consider  as a first--order 
approximation [here we include the amplitude resulting from 
the calculation with expressions (\ref{S}), (\ref{U}) approximately]:

\begin{equation}
\overline{S_t}(i)\approx 0.98S_0 \cos i
\label{Sap}
,\end{equation}
\begin{equation}
\overline{U_t}(i)\approx 1.18U_0 \cos i
\label{Uap}
.\end{equation}

To facilitate the comparison of our theory with the data, we estimate  the
correlations  of $S_t$ and $U_t$ [full expressions (\ref{S}), (\ref{U})] with the
function $\cos i$. 

\begin{equation}
\langle \overline{S_t} \cos i\rangle -\langle \overline{S_t}\rangle 
\langle \cos i\rangle=0.53S_0
\label{Sco}
,\end{equation} 
\begin{equation}
\langle \overline{U_t}\cos i\rangle -\langle \overline{U_t}\rangle\langle 
\cos i\rangle=0.60U_0
\label{Uco}
.\end{equation} 

\subsection{Amplitude of the S-component: $S_0$}
\label{.amps}

From L\'opez-Corredoira et al. (2002, Figs. 10, 11, Eq. (39)), we can
derive roughly that the maximum height $y$ of the $m=1$ component of
warp of a Milky Way-like
galaxy and baryonic mean density of the
intergalactic medium $\rho _b$ (roughly 
the average density of the IGM
flows radially ejected from the void to produce the observed effect):
\begin{equation}
y=2.8\times 10^{18}\overline{v_1} ^2({\rm km/s})
\rho _b({\rm kg/m}^3)\exp{[0.43\ x({\rm kpc})]} \ {\rm kpc}
.\end{equation}
With the definition given in Eq. (\ref{formulaW}), 
$\overline{v_1}=200$ km/s, multiplying by a factor $2/\pi $
for averaging the integration of the line of nodes over all the angles, 
the maximum amplitude is
\begin{equation}
|W_{(r\ or\ l)}|(\theta =0)=\frac{2.1\times 10^{22}\rho _b({\rm kg/m}^3)}
{L({\rm kpc})^3}
\end{equation}\[
\times \left[(\exp{[0.43L({\rm kpc})]}
[L({\rm kpc}) -2.33])+2.33\right]
.\]
The size (semiaxis length) of a Milky Way-like galaxy is approximately
$L=15$ kpc. With this number,

\begin{equation}
|W_{(r\ or\ l)}|(\theta=0)\sim 5\times 10^{22}\rho _b({\rm kg/m}^3)
\label{ampWs}
.\end{equation}
That is, due to $S=2W$,

\begin{equation} 
S_0\sim 10^{23}\rho _b({\rm kg/m}^3)
\label{s0}
.\end{equation}
We must bear in mind that
this is only an estimation of the order of magnitude, because not all the
galaxies are like the Milky Way (although this is a reasonably
good approximation for the galaxies in our sample).
To give an example, with an average  intergalactic medium  density  given by
$\langle\rho _b \rangle =\Omega _b\rho _{crit} \sim 8\times 10^{-28}$ kg/m$^3$
(taking $\Omega _b=0.042$; Spergel et al. 2007), we get a S-component with
$S_0\sim 8\times 10^{-5}$.

\subsection{Amplitude of the U-component: $U_0$}
\label{.ampu}

Similarly, from L\'opez-Corredoira et al. (2002, Figs. 10, 11, Eq. (45)), we
derive roughly that the maximum height $y$ of the $m=0$ component of the 
warp of a Milky Way-like
galaxy with IGM baryonic mean density $\rho _b$ is
\begin{equation}
y=8.2\times 10^{17}\overline{v_1} ^2({\rm km/s})
\rho _b({\rm kg/m}^3)\exp{[0.38\ x({\rm kpc})]} \ {\rm kpc}
.\end{equation}
With the definition given in (\ref{formulaW}), $\overline{v_1}=200$ km/s, 
multiplying by a factor $2/\pi $ for averaging
the integration of the line of nodes over the all angles, the
maximum amplitude is

\begin{equation}
|W_{(r\ or\ l)}|(\theta =0)=\frac{7.0\times 10^{21}\rho _b({\rm kg/m}^3)}
{L({\rm kpc})^3}
\end{equation}\[
\times \left[(\exp{[0.38L({\rm kpc})]}
[L({\rm kpc}) -2.63])+2.63\right]
.\]
With $L=15$ kpc,

\begin{equation}
|W_{(r\ or\ l)}|(\theta=0)\sim 0.8\times 10^{22}\rho _b({\rm kg/m}^3)
\label{ampWu}
;\end{equation}
that is, due to $U=2|W|$,

\begin{equation} 
U_0\sim 1.6\times 10^{22}\rho _b({\rm kg/m}^3)
\label{u0}
.\end{equation}
Again, we call
that this is a rough estimation with Milky Way-like galaxies.
With an average  intergalactic medium  density  given by
$\langle\rho _b \rangle =\Omega _b\rho _{crit} \sim 8\times 10^{-28}$ kg/m$^3$
(taking $\Omega _b=0.042$; Spergel et al. 2007), we get a U-component with
$U_0=1.3\times 10^{-5}$.

\section{Data and analysis}

We used the data of the SDSS-DR3 (3rd. data release) 
that have already been used in 
Trujillo et al. (2006). These data are their edge-on 
(inclination larger than 78$^\circ $) galaxies, which are within the shells 
$r_{void}<r<r_{void}+4\ h^{-1}$Mpc surrounding the largest voids, where 
$r_{void}>10\ h^{-1}$Mpc is its radius. 
The lower the inclination of the galaxy, the
greater the thickness of the projected disc and, consequently, the greater
the error in the determination of the centroid of $y(x)$. In the worst
case (78$^\circ $), the thickness of the projected disc is comparable 
to its intrinsic thickness (Dalcanton \& Bernstein 2002), 
so the error in the warp amplitude
is not significantly increased with respect to a 90$^\circ $
inclination galaxy.

The voids were located
using maximal spheres empty of galaxies with magnitude over 
$-19.31-5\log h$ ($H_0=100h$ km/s/Mpc), 
and they were found by means of the HB void finder
(Patiri et al. 2006). From the SDSS available public data,
we used the filter ``r'' images. In total we have 114 galaxies.

\begin{figure*}
\vspace{1cm}
{\par\centering \resizebox*{15cm}{5cm}{\includegraphics{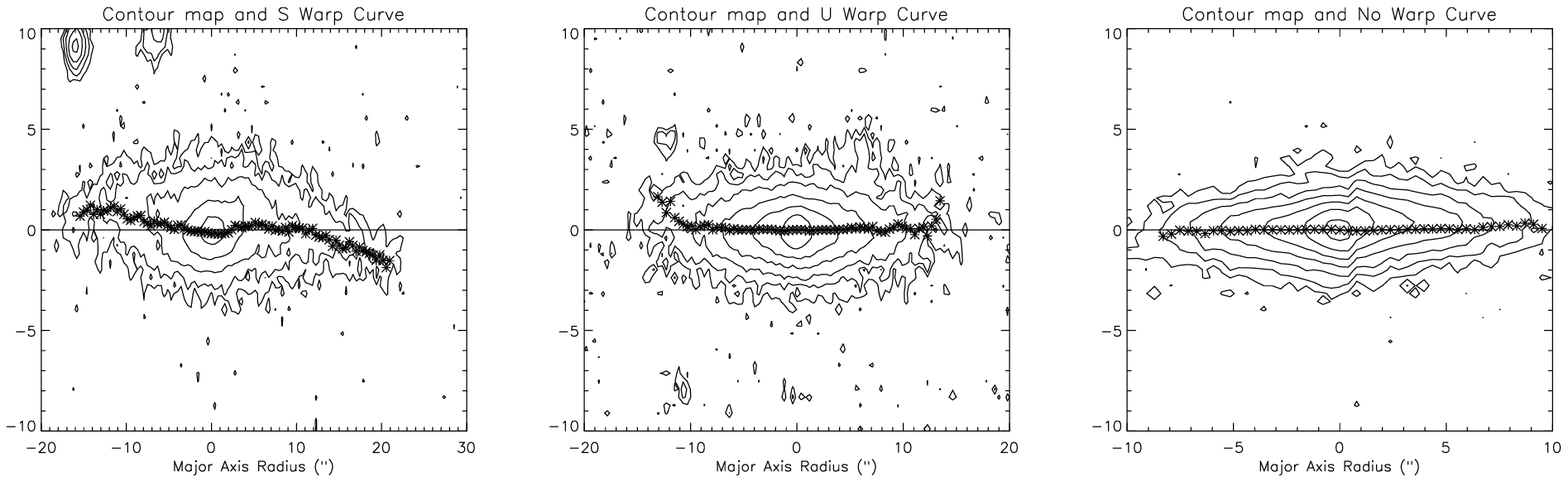}}\par
\hspace{4mm}\resizebox*{4cm}{4cm}{\includegraphics{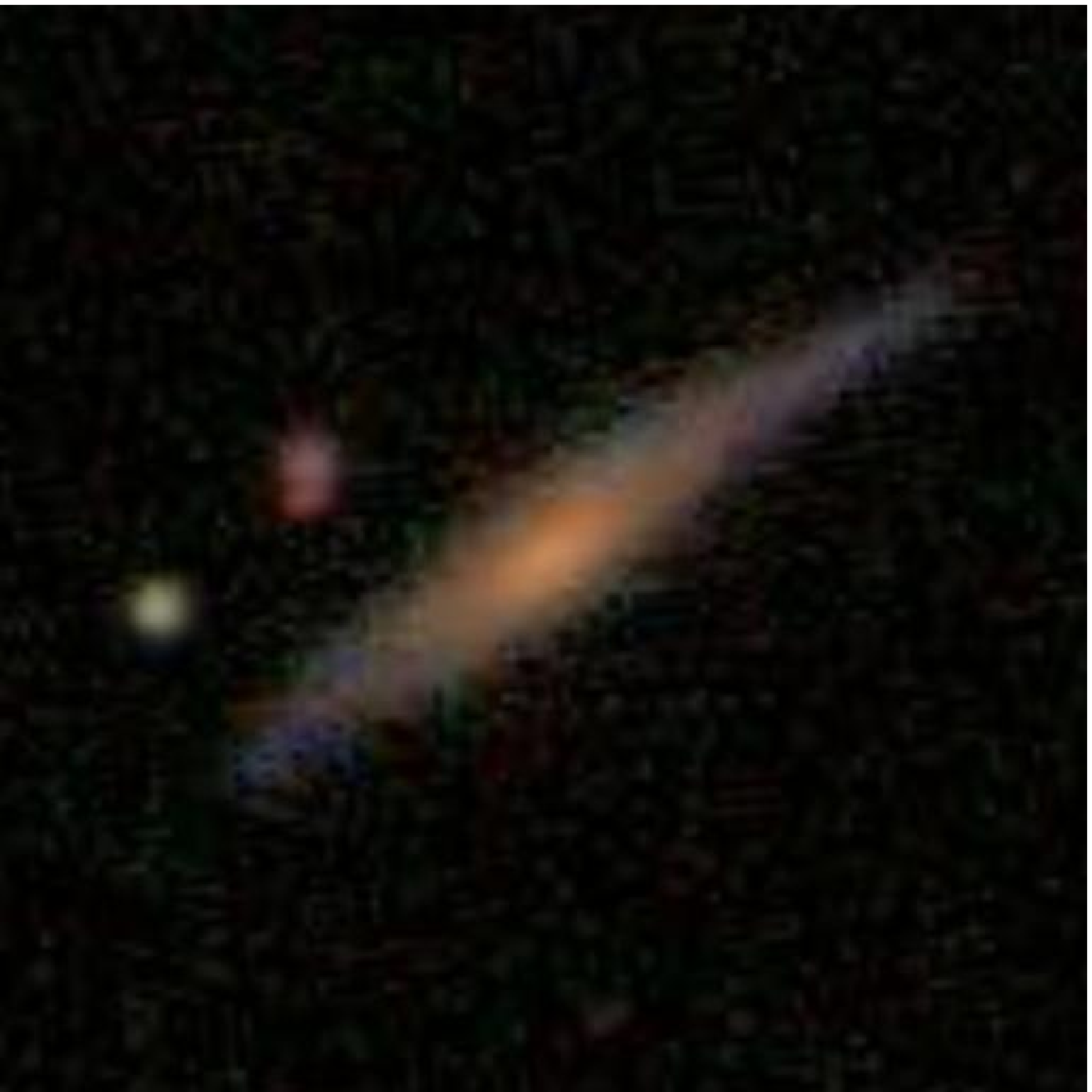}}\hspace{4mm}
\resizebox*{4cm}{4cm}{\includegraphics{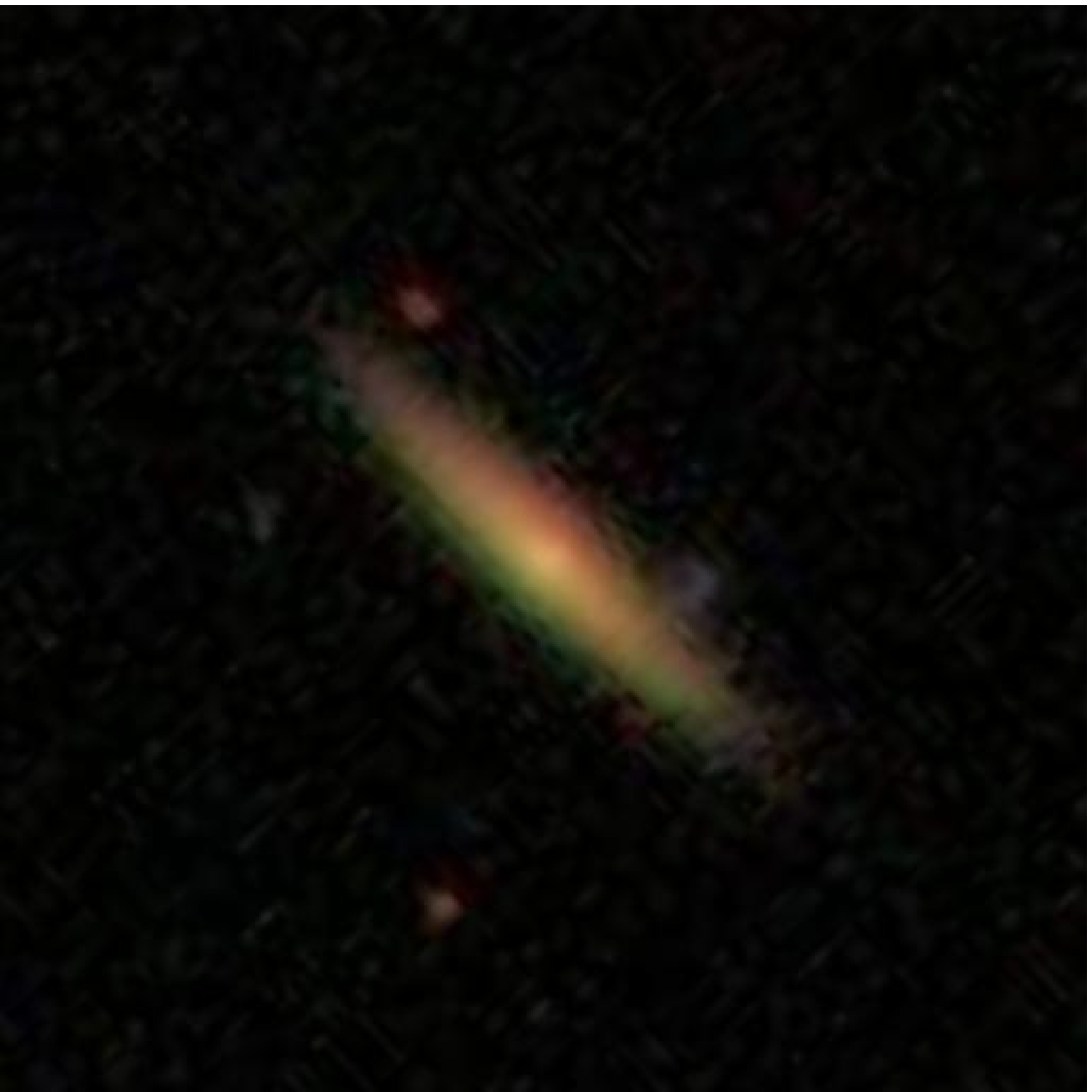}}\hspace{4mm}
\resizebox*{4cm}{4cm}{\includegraphics{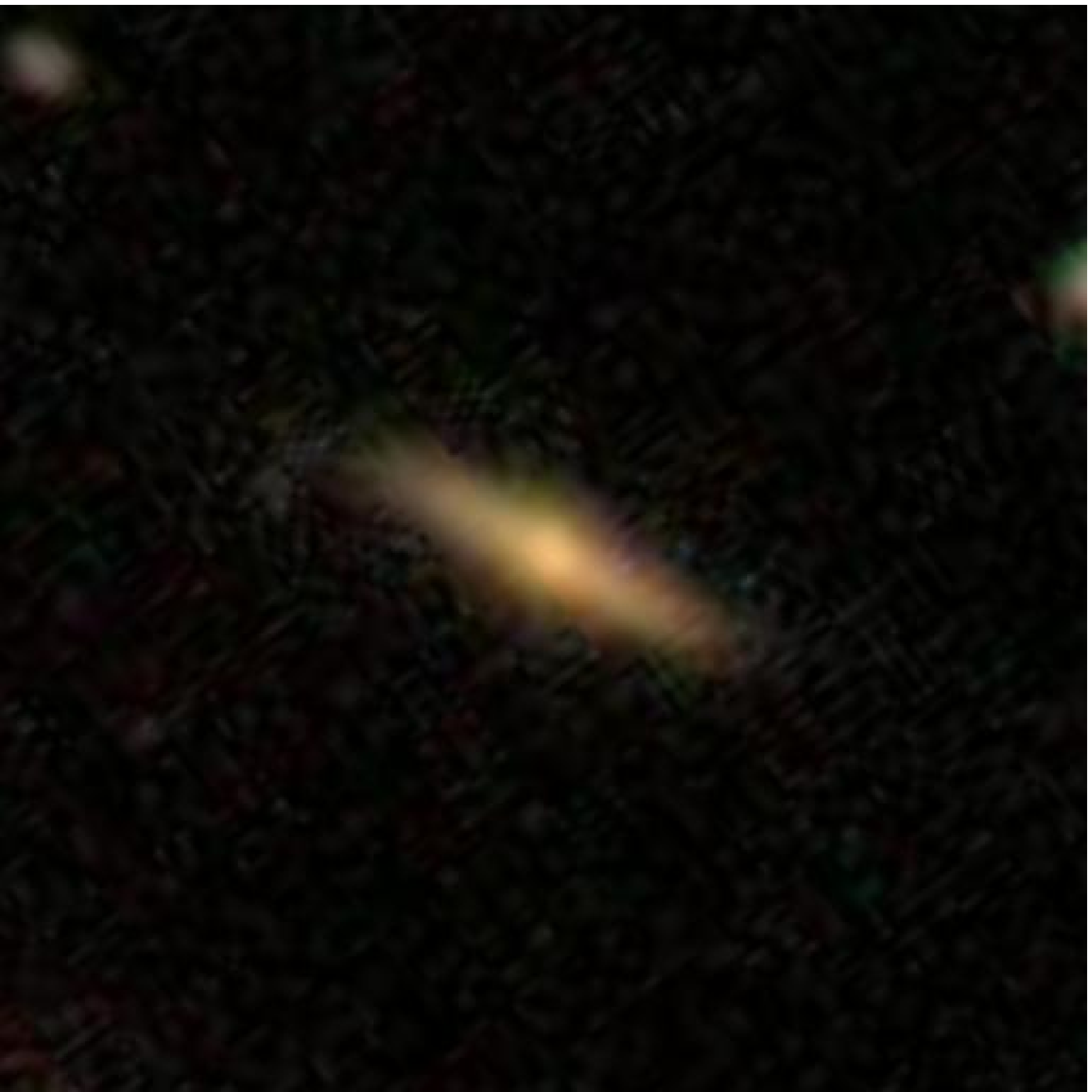}} \par}
\caption{Warp curves and contour maps of three selected galaxies (up);
and 5-filters combined SDSS images of them (down), 50"x50".
The isophotes are equidistant (in units of n*3$\sigma$ 
equiv. to a step of +0.75 mag/arcsec$^2$) starting at a level of about 3
$\sigma$ above the sky background.
Left: $R.A.=234.8898^\circ $, $\delta =+3.2352^\circ $ (J2000),
S-warped. Centre: $R.A.=131.4264^\circ $, $\delta =+51.2122^\circ $
(J2000), U-warped downwards. Right: $R.A.=226.3795^\circ $, 
$\delta =+57.4142^\circ $ (J2000), negligible warp.}
\label{Fig:sample}
\end{figure*}

For seventeen galaxies there were difficulties measuring the warp amplitude
(for instance, due to the proximity of a star in the field or interaction
with other galaxies), so there
remain $N=97$ galaxies with which we carried out the statistics
(Table \ref{Tab:data}). In Fig. \ref{Fig:sample}, we
show three examples.

Some images of warped galaxies could be the subject of alternative
interpretations. For instance, considering the isophote maps, warp curve,
and image in the central panel of Fig, \ref{Fig:sample}, 
a feature is found at
$x=-12$, $y=5$, either a companion galaxy or an inteloper, 
which could produce/modify the warp curve. However, we find that 
the warp at this galactocentric radius is real, as directly deduced 
from a detailed study of the isophote maps.

\begin{figure*}
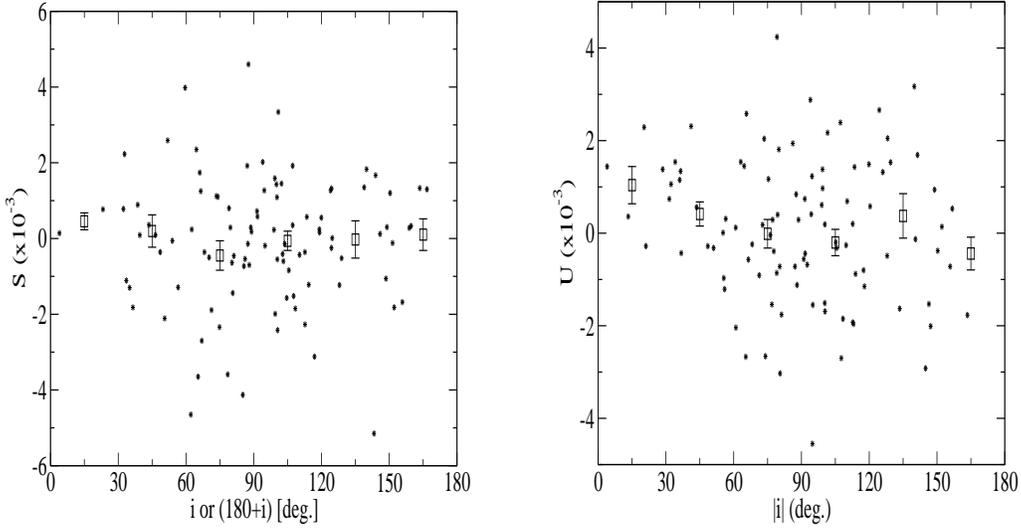

\vspace{1cm}
{\par\centering \resizebox*{6.2cm}{7cm}{\includegraphics{fort50.eps}}
\hspace{1cm}\resizebox*{6.2cm}{7cm}{\includegraphics{fort51.eps}}\par}
\caption{Dependence of $S$ and $U$ on the inclination $i$ 
in the observational data (stars). The squares with error bars
represent the average in bins of $i$ of 30 degrees.}
\label{Fig:stat}
\end{figure*}

If we plot their values of $S$ and $U$ vs. $i$, we get the
results of Fig. \ref{Fig:stat}.
There are slight trends in $S(i)$ [$\langle S\ i(^\circ )\rangle
-\langle S\rangle\langle i(^\circ )\rangle=
(0.3^\circ\pm 6.1^\circ )\times 10^{-3}$] and in $U(i)$ 
[$\langle U\ i(^\circ )\rangle
-\langle U\rangle\langle i(^\circ )\rangle=
(-8.6^\circ\pm 5.6^\circ )\times 10^{-3}$].
The errors in the correlations are calculated as $\sigma _S\sigma _i/\sqrt{N}$
and $\sigma _U\sigma _i/\sqrt{N}$; where $\sigma _S$, $\sigma _U$
and $\sigma _i$ are the r.m.s. of the values of $S$, $U$, and $i$.

The scattering of Fig. \ref{Fig:stat} may be for several
reasons. For example,

\begin{enumerate}

\item The measurement of the warp amplitude has errors.

\item Several mechanisms produce warps. The accretion produces
signal and noise, while the other mechanisms only produce noise.

\item Different masses of the galaxies not taken into account 
in our model, as said in \S \ref{.amps} and \ref{.ampu}). 

\item The wind's other components apart from the radial one introduce
scattering

\item The error of the inclination with respect to the void also
introduces scattering.

\end{enumerate} 
 
All these sources of contamination introduce an increase in the 
scattering but not a systematic error. We can consider the error of 
each individual point in Fig. \ref{Fig:stat}
as $\sigma _S=1.7\times 10^{-3}$ for the S-component and $\sigma _U=1.6\times
10^{-3}$ for the U-component (r.m.s. in Fig. \ref{Fig:stat}). 
Our mission is to extract the statistical information hidden 
behind these clouds of points.

\subsection{Checking the null-hypothesis}

To check whether the null hypothesis is compatible with the observed
U-components distribution, 
we computed the probability, $P(i_*)$, based on the binomial distribution
of finding no more than $n_0^-$ galaxies with $U<0$ and
$|i|\le i_*$, and no more than $n_1^+$ galaxies with $U>0$ and
$|i|\ge 180-i_*$, assuming that there is no correlation between $i$
and $U$ (i.e., the null-hypothesis). With this assumption, 
the probability that $U>0$ is 1/2 for
any value of $i$, and using the binomial distribution for 
$n_1^+$, $n_0^-$, we find

\begin{equation}
P(i_*)=\left(\sum _{i=0}^{n^¯_0}\left(\begin{array}{l}
m_0 \\ 
i \end{array}\right)\right)
\left(\sum _{i=0}^{n_1^+}\left(\begin{array}{l}
m_1 \\ 
i \end{array}\right)\right)2^{-(m_0+m_1)}
,\end{equation}
where $m_0$ is the number of galaxies with $|i|\le i_*$;
$m_1$ is the number of galaxies with $|i|\ge (180-i_*)$.
To determine $i_*$, we assume that the signal is proportional
to $\cos i$, as determined previously in our model [see Eq. (\ref{Uap})].
We assume that the galaxies are uniformly distributed in $i$ (the small alignment reported
by Trujillo et al. 2006 is not very relevant to this purpose), then
\begin{equation}
\langle \overline{U}\rangle(i_*)=\frac{1.18U_0 \int _0^{i_*} \sin i
\cos i\ di}{\int _0^{i_*} \sin i\ di}\propto (1+\cos {i_*})
\label{Uaver}
.\end{equation}
The r.m.s. is proportional to the number of galaxies within $i<i_*$,
which for a given sample and again assuming isotropy, has the proportionality
\begin{equation}
\sigma _{\overline{U}}(i_*)\propto (1-\cos {i_*})^{-1/2}
.\end{equation}
Thus, according to our model, the value of $i_*$ that maximizes the signal--to
--noise ratio, $\frac{\langle \overline{U}\rangle}
{\sigma _{\overline{U}}}$, is given for $i_*=60^\circ $.
If instead of the approximate cosine dependence of Eq. (\ref{Uap}),
we took the exact calculation of Eq. (\ref{U}), the value that
maximizes the signal to noise ratio would be $i_*=73^\circ $ for
U-component; for S-component, it would be $i_*=75^\circ $.
We assume as a good approximation the cosine dependence, so we
use $i_*=60^\circ $.

With this value of $i_*=60^\circ $, the probability that our data are
compatible with the null hypothesis is
$P=0.056$; that is, the null hypothesis is excluded
within 94.4\% C.L. If we took $i_*=45^\circ $, we would get
$P=0.0043$ (rejection of non-correlation within 99.57\% C.L.), but
this value of $i_*$ is not justified, a priori; therefore, the statistical
significance must be less than this. For a higher value of $i_*$,
we also get rejection of the null hypothesis.
For $i_*=75^\circ $ we get rejection within 95.7\% C.L.

If we do the same calculation for the $S$ vs. $i$ data, we find
that the probability of null hypothesis cannot be rejected 
($P(i_*=60^\circ )=0.13$). 
We also checked the null hypothesis with 
the Spearman rank correlation coefficient. This test
gives higher probabilities of a null correlation:
$P=0.154$ for $U(i)$ and $P=0.739$ for $S(i)$.

\subsection{Most likely values of the signal}

If we assume that Eq. (\ref{Uap}) with some positive $U_0$ applies,
the mean value of $U$ for $i\le i_*$, $\overline{U}$, is given
by Eq. (\ref{Uaver}):
\begin{equation}
\overline{U}=0.88U_0
\label{u0uav}
.\end{equation}
The most likely value of $\overline{U}$ can be estimated
from the data as follows.

We assume that the distribution
of the probabilities of a value of $U$, $P(U)$, is Gaussian, centred
at $\overline{U}$ with r.m.s. $\sigma _U$.
For galaxies with $i\le i_*$, the probability that $U<0$ is 
\begin{equation}
\int _{-\infty }^0P(U)dU\equiv \frac{1}{2}-\omega 
;\end{equation}
i.e.,
\begin{equation}
\omega (U)=\frac{1}{2}\left(1- erfc\left(\frac{U}{\sqrt{2}\sigma _U}
\right)\right)
,\end{equation}
and the probability that $U>0$ is 
\begin{equation}
\int _0^\infty P(U)dU=\frac{1}{2}+\omega 
.\end{equation}
Identically, for $i\ge 180-i_*$ the probability 
that $U<0$ is $\frac{1}{2}+\omega $, and the probability that $U>0$
is $\frac{1}{2}-\omega $. The probability, $F(\omega )$, that the
number of galaxies with $i\le i_*$ and $U<0$, $k$, 
is smaller than or equal to the
observed value $n_0^-$ (from a total of $m_0$ galaxies with $i\le i_*$)
and that, for $i\ge 180-i_*$ the number of galaxies with $U>0$, $j$, is
lower than or equal to the observed value, $n_1^+$ 
(from a total of $m_1$ galaxies with $i\ge 180-i_*$), is given by
the multiplication of the probabilities of both events:

\begin{equation}
F(\omega)=\left(\sum _{k=0}^{n_0^¯}\left(\begin{array}{l}
m_0 \\ 
k \end{array}\right)(1/2+\omega)^{m_0-k}
(1/2-\omega)^{k}\right)
\end{equation}\[\times
\left(\sum _{j=0}^{n_1^+}\left(\begin{array}{l}
m_1 \\ 
j \end{array}\right)(1/2+\omega)^{m_1-j}
(1/2-\omega)^{j}\right)
.\]
Thus, the mean value $\overline{U}$ 
is obtained from $F[\omega (\overline{U})]=0.5$,
and the maximum and minimum values within 95\% C.L.
would be respectively $\overline{U}_+$, $\overline{U}_-$ 
derived from $F[\omega (\overline{U}_+)]=
0.95$, $F[\omega (\overline{U}_-)]=0.05$ respectively. 
Note that $\omega (U)$ has been
defined so that it must be positive if our model applies. A negative
value of $\omega (U)$ should be interpreted as evidence against it.

With our data and $i_*=60^\circ $, the values are: 
$\overline{U}=6.7\times 10^{-4}$, $\overline{U}_+=14.5\times
10^{-4}$, $\overline{U}_-=-0.3\times 10^{-4}$.
Using our model, we can use these numbers to put a constraint
on the IGM density.
From Eqs. (\ref{u0uav}) and (\ref{u0}), we find
that $\langle \rho _b\rangle =4.8^{+5.5}_{-4.9} \times 10^{-26}$ kg/m$^3$ 
(95\% C.L.). This allows us to put an upper limit on the IGM density 
but not a minimum. 
The same calculation with $S$-component gives a tighter constrain:
$\overline{S}=5.1\times 10^{-4}$, $\overline{S}_+=13.6\times
10^{-4}$, $\overline{S}_-=-2.3\times 10^{-4}$;
$\langle \rho _b\rangle =6.7^{+11.4}_{-9.8} 
\times 10^{-27}$ kg/m$^3$ (95\% C.L.).

The correlation of $S$ with a cosine function
(approximately the expected shape theoretically) is 
$\langle S\ \cos i \rangle
-\langle S\rangle\langle \cos i\rangle=(-0.004\pm 0.091)\times 10^{-3}$.  
The correlation of $U$ with $U_t$ 
(also approximated to be a cosine function) is $\langle U\ \cos i\rangle
-\langle U\rangle\langle \cos i\rangle=
(0.116\pm 0.083)\times 10^{-3}$.
We can get a better constraint for the maximum density from
these correlations: with the S-component measurement 
and the expressions (\ref{Sco}) and (\ref{s0}),
we find that $\langle \rho _b\rangle < \sim 3\times 10^{-27}$ kg/m$^3\approx
4\Omega _b\rho _{crit}$ (95\% C.L.$\equiv 2\sigma $);
with the U-component measurement and the expressions (\ref{Uco}) and (\ref{u0}),
we find that $\langle \rho _b\rangle < 3\sim 10^{-26}$ kg/m$^3\approx
37\Omega _b\rho _{crit}$ (95\% C.L.).

Therefore, summarising the contents of this section, 
we reject the null hypothesis (i.e., the inclination of galaxies
and the amplitude of the warp are not related to each other) at
94.4\% C.L. Using our model, we can estimate the average density of
the radial flow from the void to be
0-4$\left(\frac{\overline{v_1}}{200\ {\rm km/s}}\right)^{-2}
\Omega _b\rho _{crit}$.

\section{Conclusions}

Cosmological hydrodynamical simulations predict flows of IGM along the radial
vector of the void. This radial direction is approximately the same as the
infall of matter in the early stages of the galaxy formation at the shells of
the void. One way to search for the effect of this IGM flow 
in these shells is to measure the
dependence of the warp amplitude on their galaxies as a function of
their inclination with respect to the radial vector of the void. In
this paper, we have developed a method to measure that effect, and
we made a first attempt to find this effect.
The signal found in the U component of the warp (the null hypothesis is 
rejected at 94.4\% C.L.) gives some hint
that such an effect might exist. This result is not 
conclusive (5.6\% is not a very negligible probability)
and the absence of the radial flows cannot be excluded at present.
If the IGM radial flows in the radial direction of the voids exist, 
their baryonic matter density should be 
$\langle \rho _b\rangle  <\sim 3\times 10^{-27}$
kg/m$^3=4\Omega _b\rho _{crit}$. This density would increase inversely
proportional to the square of the mean flow velocity if its value differs from
200 km/s. There is also the possibility that the accretion of material have
different initial velocities than the radial direction of the void. 

There may be other mechanisms of warp formation different to
the accretion onto the disc, but they would produce noise in the
correlation if they have nothing to do with the IGM accretion.
If the correlation of S-component amplitude and inclination were observed, 
although it would be an argument
in favour of L\'opez-Corredoira et al. (2002) theory, it would not
be totally conclusive because there might be alternative explanations 
for the correlation. The mechanism of accretion into the halo (Ostriker \& Binney
1989; Jiang \& Binney 1999) rather than onto the disc might possibly
explain the correlation. There might be a relationship 
between warps and filaments associated to the void
produced by primordial magnetic fields, or the frozen magnetic 
fields were aligned with the filaments (Florido \& Battaner 1997), 
if the magnetic fields are also responsible for the warp formation 
(Battaner et al. 1990; Battaner et al. 1991; 
Battaner \& Jim\'enez--Vicente 1998).
However, these theories do not explain the U-component (the asymmetry
of the S-warps), which are clearly observed in many galaxies (e.g., 
Reshetnikov \& Combes 1998; S\'anchez-Saavedra et al. 2003). 
The trend in the correlation of the U-component with the inclination of
the galaxy obtained in this paper, if confirmed with higher statistical 
significance, could be taken as confirmation that the mechanism of 
IGM accretion onto the disc produces warps. 
The application of the method presented in this paper to 
galaxy samples with more objects and/or better measurements
of the warp amplitude is expected to give more accurate results.

\

{\bf Acknowledgments:}
Thanks are given to the anonymous referee for helpful comments,
and to Joly Adams (language editor of A\&A) for proof-reading this
paper. Funding for the creation and distribution of the SDSS Archive has been
provided by the Alfred P. Sloan Foundation, the Participating  Institutions, the
National Aeronautics and Space Administration, the  National Science Foundation,
the U.S. Department of Energy, the Japanese Monbukagakusho, and the Max Planck
Society. The SDSS Web site is  http://www.sdss.org/. The SDSS is managed by the
Astrophysical Research  Consortium (ARC) for the Participating Institutions. The
Participating  Institutions are The University of Chicago, Fermilab, the
Institute for  Advanced Study, the Japan Participation Group, The Johns Hopkins
University, the Korean Scientist Group, Los Alamos National Laboratory, the
Max-Planck-Institute for Astronomy (MPIA), the Max-Planck-Institute for 
Astrophysics (MPA), New Mexico State University, University of Pittsburgh, 
University of Portsmouth, Princeton University, the United States Naval 
Observatory, and the University of Washington.
MLC was supported by the {\it Ram\'on y Cajal} Programme
of the Spanish Science Ministery.
We thank the Spanish Science Ministery for support under
grant AYA2007-67625-CO2-01.

\begin{table}
\caption{Amplitude of the warp (with the corresponding error), in
units of $10^{-3}$, and
inclination with respect to the radial direction of the void
of the used SDSS galaxies in this paper.}
\label{Tab:data}
\begin{center}
\begin{tabular}{llllll}
R.A.($^\circ $) & Decl.($^\circ $) & $W_r$  &  $W_l$ & 
$Err(W_{r/l})$ & $i(^\circ )$ \\ \hline
  7.2201 & -10.1462 & -0.8 & -2.3 &  1.4 &  -63.1 \\  
  8.1775 & -10.7132 &  0.1 &  0.5 &  1.7 &   91.6 \\  
 12.7354 & -10.3630 & -0.6 &  0.4 &  1.5 &  -55.6 \\ 
 16.2514 & -10.6816 & -0.4 & -0.1 &  2.5 &  -51.1 \\   
 37.5906 &  -1.1753 &  0.0 &  0.1 &  0.9 & -140.5 \\  
 44.4492 &  -7.7039 & -0.8 &  0.9 &  1.5 & -133.5 \\  
 52.2561 &  -0.2598 &  1.3 & -0.1 &  3.2 & -113.5 \\  
116.3311 &  33.9423 &  0.0 &  0.8 &  0.7 &   79.1 \\  
117.2566 &  40.5368 & -2.1 &  0.6 &  1.6 &  107.6 \\  
118.8881 &  43.4200 &  2.3 &  1.7 &  2.3 & -120.4 \\  
119.9286 &  44.3550 &  0.1 & -0.6 &  1.8 &  110.3 \\  
120.5069 &  26.0021 &  0.5 &  0.9 &  2.4 &  -77.7 \\  
123.8789 &  41.7845 &  2.5 & -0.4 &  2.1 &   94.0 \\  
126.8650 &  34.0610 & -1.0 & -1.4 &  2.4 &  -79.5 \\  
127.0763 &  30.4663 &  2.7 &  1.9 &  1.5 &   87.6 \\  
127.3464 &  32.3032 & -0.1 & -0.3 &  0.5 &  112.7 \\  
127.3562 &  34.7077 &  1.0 &  0.9 &  1.7 &  -72.8 \\  
128.1052 &  43.3659 & -2.3 & -0.4 &  1.4 & -113.0 \\  
129.5621 &  46.6003 &  0.4 & -0.2 &  1.1 &   99.0 \\  
129.9943 &  53.1409 & -0.9 &  0.7 &  1.8 &   95.0 \\  
130.3608 &  43.7136 & -2.1 &  0.8 &  2.2 & -144.9 \\  
131.4264 &  51.2122 &  0.7 & -1.9 &  0.7 &  -65.6 \\  
131.9592 &  46.4286 &  0.7 &  0.0 &  1.4 &   91.5 \\  
133.3355 &  37.7273 & -2.5 &  0.5 &  3.9 &  -80.5 \\  
139.2459 &   3.2982 & -0.1 & -0.3 &  1.1 &  -77.2 \\  
139.4740 &  57.2313 & -0.2 & -1.6 &  1.5 &   36.5 \\  
140.1681 &  46.3610 & -0.3 & -1.8 &  3.0 & -129.4 \\  
140.9270 &  56.9905 &  2.0 & -0.7 &  0.8 &  124.4 \\  
141.6051 &  46.8787 &  0.6 &  0.1 &  3.8 & -156.8 \\  
143.5772 &  48.0993 &  0.8 & -0.6 &  1.1 &    4.0 \\  
145.8396 &   1.7637 &  1.9 &  0.4 &  2.0 &   64.6 \\  
148.3098 &   8.9179 &  0.5 & -0.1 &  1.1 &   43.6 \\  
151.3779 &   5.6446 &  0.1 &  2.1 &  1.6 & -147.2 \\  
151.6732 &   5.4257 & -0.2 &  1.3 &  1.9 &  100.3 \\  
161.8397 &  65.1576 & -1.1 &  0.6 &  1.5 &  100.4 \\  
162.2562 &  59.4725 & -0.7 & -2.9 &  0.9 & -101.6 \\  
%
163.8244 &  48.8133 &  2.3 &  0.3 &  1.0 & -128.1 \\  
164.7096 &   0.7530 &  1.4 & -1.0 &  2.8 &  107.2 \\  
165.8762 &  50.9933 &  1.2 &  0.0 &  3.0 &   94.7 \\  
167.3524 &  65.7939 &  0.9 & -0.1 &  0.9 &   32.3 \\  
168.3881 &  58.0182 &  1.6 & -0.5 &  2.4 &   73.5 \\  
172.3922 &  57.3641 & -2.8 & -2.4 &  1.8 &  -36.8 \\  
175.2166 &  -2.8630 & -0.7 &  0.0 &  2.5 &   80.4 \\  
176.9355 &  62.8054 & -0.8 & -1.0 &  0.7 &  152.2 \\  
178.7274 &   1.3727 & -0.2 & -1.4 &  0.9 &  -75.4 \\  
180.3058 &  56.1816 & -0.5 & -0.8 &  0.8 &   56.5 \\  
181.1288 &  54.6058 & -0.2 &  0.4 &  1.4 &  -91.0 \\  
181.5367 &   4.5987 &  0.4 &  1.3 &  1.2 & -113.9 \\  
181.8765 &  -2.0978 & -0.4 & -0.1 &  2.4 & -109.8 \\  
185.1264 &  -1.4101 & -0.3 &  0.5 &  2.1 & -117.5 \\  
187.4005 &  -3.7120 &  0.8 & -0.7 &  2.3 &  -34.1 \\  
187.4013 &  -1.0802 &  0.4 &  0.8 &  1.3 &  150.3 \\  
189.8933 &   5.0846 & -3.2 & -0.5 &  1.0 &   65.4 \\  
191.5689 &  62.2786 & -0.9 &  0.2 &  1.7 &   88.0 \\  
193.1906 &  62.6491 & -1.1 &  0.6 &  1.5 &   81.2 \\  
\end{tabular}
\end{center}
\end{table}   

\begin{table}
Cont. table \protect{\ref{Tab:data}}.
\begin{center}
\begin{tabular}{llllll}    
R.A. & Decl. & $W_r$   &  $W_l$ &
$Err(W_{r/l})$ & $i(^\circ )$ \\ \hline 
195.1273 &   0.4801 & -0.4 &  0.3 &  0.5 &  -92.6 \\  
195.7355 &   3.7370 & -1.3 &  0.2 &  1.4 & -146.4 \\  
195.7591 &  -2.4564 & -0.6 & -0.3 &  1.1 &  105.5 \\  
197.1058 &  61.4598 &  0.6 & -0.7 &  0.9 & -126.0 \\  
197.8032 &  49.4153 &  0.0 & -1.4 &  0.9 &  -99.3 \\  
198.0996 &  -2.8867 &  1.6 & -0.2 &  1.4 &  -80.0 \\  
198.5669 &  -2.0749 &  0.1 &  0.0 &  1.7 &  -60.9 \\  
198.9388 &  59.8330 &  0.8 &  0.5 &  1.1 &  -13.3 \\  
199.0415 &  65.9117 & -4.3 &  0.2 &  2.3 &  -94.9 \\  
204.4937 &   5.7839 &  0.6 &  1.3 &  2.3 &   87.1 \\  
207.0692 &  46.7701 & -1.2 & -0.5 &  1.1 &  155.8 \\  
207.7140 &   5.2116 & -0.9 & -0.4 &  1.3 &  128.0 \\  
212.5481 &  58.7175 &  1.8 & -0.5 &  1.9 &  -41.1 \\  
214.8475 &   2.8468 &  2.5 & -0.7 &  1.5 &  139.9 \\  
219.3559 &   0.4227 &  0.2 &  0.1 &  0.5 & -100.3 \\  
220.0357 &   3.0825 & -0.3 & -0.1 &  0.3 &   68.2 \\  
220.3983 &  62.7731 &  3.8 & -0.4 &  2.8 &  -79.1 \\  
220.7501 &  61.6237 &  1.0 & -0.5 &  1.6 &  119.9 \\  
223.3199 &  49.1950 & -0.2 &  1.5 &  1.5 &  163.5 \\  
225.0443 &  43.1852 &  0.6 & -0.3 &  1.6 &  148.9 \\  
226.3795 &  57.4142 &  1.3 &  0.3 &  1.2 &   99.4 \\  
226.8317 &   0.5954 & -1.3 & -1.1 &  1.1 & -105.2 \\  
227.2538 &  41.3454 & -1.4 & -0.5 &  1.6 &   71.3 \\  
230.8046 &  39.8048 &  0.7 & -1.2 &  1.5 &   86.2 \\  
233.1236 &  50.0678 & -0.8 &  1.9 &  1.7 &   74.1 \\  
233.7380 &  58.4997 & -0.3 &  0.0 &  0.8 &   48.6 \\  
234.8898 &   3.2352 & -2.9 & -1.8 &  2.4 & -117.9 \\  
236.9465 &  55.5457 &  0.6 & -0.8 &  1.5 &  -28.5 \\  
238.5349 &  45.0029 &  0.0 &  0.3 &  1.3 &  -21.1 \\  
239.2935 &  45.2622 &  0.3 &  0.0 &  0.7 &   88.7 \\  
241.3029 &  42.8759 & -0.9 &  1.1 &  1.5 &  -61.0 \\  
242.0378 &  50.1382 & -0.1 & -0.1 &  1.2 &  -76.3 \\  
243.6710 &  38.0918 & -0.2 & -0.9 &  1.6 &  -31.5 \\  
244.8777 &  42.4485 &  1.3 & -1.0 &  1.7 &  -20.4 \\  
249.2328 &  44.0929 &  1.4 &  0.3 &  2.1 &  -36.1 \\  
251.6845 &  42.7894 &  0.0 &  1.2 &  1.9 &  -56.0 \\  
254.6449 &  32.6595 &  0.0 &  0.6 &  0.7 &  -66.5 \\  
257.1461 &  59.3407 & -0.2 & -0.6 &  0.7 &  -94.4 \\  
258.1745 &  30.4143 & -2.1 & -0.2 &  3.2 &  112.7 \\  
324.0977 &  -6.4763 & -1.9 &  0.0 &  0.8 &  108.4 \\  
341.0004 &  -0.9310 &  1.3 & -0.4 &  3.0 & -141.4 \\  
354.9044 &  14.5762 & -1.1 &  0.5 &  1.3 &  -76.9 \\  
\end{tabular}
\end{center}
\end{table}


\begin{thebibliography}{99}

\bibitem{} Arag\'on-Calvo, M. A., van de Weygaert, R., Jones, B. J. T., 
\& van der Hulst, J. M. 2007, ApJ, 655, L5

\bibitem{} Bailin, J., \& Steinmetz, M.  2005, ApJ, 627, 647

\bibitem{} Battaner, E., Florido, E., Sanchez-Saavedra, M. L. 1990, 
A\&A, 236, 1

\bibitem{} Battaner, E., Garrido, J. L., S\'anchez-Saavedra, M. L., 
\& Florido, E. 1991, A\&A, 251, 402

\bibitem{} Battaner, E., \& Jimenez-Vicente, J. 1998, A\&A, 332, 809

\bibitem{} Betancort-Rijo, J., \& Trujillo, I. 2008, in preparation

\bibitem{} Bertin, G., \& Mark, J. W.-K. 1980, A\&A, 88, 289

\bibitem{} Binney, J., 1981, MNRAS, 196, 455

\bibitem{} Binney, J., Jiang, I.-G., \& Dutta, S. 1998, MNRAS, 297, 1237

\bibitem{} Brunino, R., Trujillo, I., Pearce, F. R., \& Thomas, P. A. 
2007, MNRAS, 375, 184

\bibitem{}
Dalcanton, J. J., \& Bernstein, R. A. 2002, AJ, 124, 1328

\bibitem{} Dekel, A., \& Shlosman, I. 1983, in: Intern. kinemat. \& Dynam.
galaxies (IAU Sym. 100), p.\ 187

\bibitem{} Florido, E., \& Battaner, E. 1997, A\&A, 327, 1

\bibitem{} Fraternali, F., Binney, J., Oosterloo, T., \& Sancisi, R.
2007, New Astron. Rev. 51, 95

\bibitem{} Garc\'\i a-Ruiz, I., Kuijken, K., \& Dubinski, J. 2002, 
MNRAS, 337, 459 

\bibitem{} Guijarro, A., Peletier, R., Battaner, E., Jimenez-Vicente, J., 
de Grijs, R., \& Florido, E. 2008, in preparation

\bibitem{} Hahn, O., Porciani, C., Carollo, C. M., \& Dekel, A. 2007,
MNRAS, 375, 489

\bibitem{} Hunter, C., \& Toomre, A. 1969, ApJ, 155, 747

\bibitem{} Jiang, I.-G., \& Binney, J. 1999, MNRAS, 303, 7

\bibitem{} Kashikawa, N., \& Okamura, S. 1992, PASJ, 44, 493

\bibitem{} Kuijken, K. 1991, ApJ, 376, 467

\bibitem{} L\'opez-Corredoira, M., Betancort-Rijo, J. E., \& Beckman,
J. E. 2002, A\&A 386, 169 

\bibitem{} Mayor, M., \& Vigroux, L. 1981, A\&A 98, 1

\bibitem{} Navarro, J. F., Abadi, M. G., \& Steinmetz, M. 2004, ApJ, 
613, L41

\bibitem{} Nelson, R. W., \& Tremaine, S. 1995, MNRAS, 275, 897

\bibitem{} Ostriker, E. C., \& Binney, J. J. 1989, MNRAS, 237, 785

\bibitem{} Patiri, S., Cuesta, A. J., Prada, F., Betancort-Rijo, J.,
\& Klypin, A. 2006, ApJ, 652, L75

\bibitem{} Paz, D., Stasyszyn, F., \& Padilla, N. 2008,
arXiv:0804.4477

\bibitem{} Porciani, C., Dekel, A., \& Hoffman, Y. 2002, 
MNRAS, 332, 339

\bibitem{} Quinn, T., \& Binney, J. 1992, MNRAS, 255, 729

\bibitem{} Reshetnikov, V., Battaner, E., Combes, F., \& 
Jim\'enez-Vicente, J. 2002, A\&A, 382, 513

\bibitem{} Reshetnikov V., \& Combes F. 1998, A\&A, 337, 9

\bibitem{} Revaz Y., \& Pfenninger D., 2001, in:
Gas and Galaxy Evolution (ASP Conf. Proc. 240), 
J. E. Hibbard, M. Rupen, J. H. van Gorkom (eds.), 
San Francisco: Astronomical Society of the Pacific, p.\ 278

\bibitem{} Saha, K., \& Jog, C. J., 2006, A\&A, 446, 897

\bibitem{} S\'anchez-Saavedra, M. L., Battaner, E, \& Florido, E.
1990, MNRAS 246, 458

\bibitem{} S\'anchez-Saavedra, M. L., Battaner, E., Guijarro, A.,
L\'opez-Corredoira, M., \& Castro-Rodr\'\i guez, N. 2003, A\&A 399,
457

\bibitem{} S\'anchez-Salcedo, F. 2006, MNRAS 365, 555

\bibitem{} Sparke, L. S., \& Casertano, S. 1988, MNRAS, 234, 873 

\bibitem{} Spergel, D. N., Bean, R., Dore, O., et al. 2007, 
ApJS 170, 377

\bibitem{} Toomre, A. 1983, in: Internal kinematics and dynamics of galaxies, 
D. Reidel Publishing Co., Dordrecht, p.\ 177

\bibitem{} Trujillo, I., Carretero, C., \& Patiri, S. 2006,
ApJ 640, L111

\bibitem{} van der Kruit, P. C. 2007, A\&A 466, 883

\bibitem{} Weinberg, M. D. 1998, MNRAS, 299, 499

\bibitem{} Zurita, A., \& Battaner, E. 1997, A\&A 322, 86

\end{thebibliography}
\end{document}